\documentclass[aps,pra,preprint,onecolumn,letterpaper]{revtex4}

\usepackage{graphicx}   
\usepackage{import}                         
\usepackage{epstopdf}
\usepackage{amsmath} 
\usepackage{bm}
\usepackage{amssymb}
\usepackage{quotes}
\usepackage{color}
\usepackage{transparent}
\usepackage{dcolumn}

\usepackage{multirow}
\usepackage{cancel} 
\usepackage{mdframed}
\usepackage{color}
\usepackage{bm}
\usepackage{dsfont}
\usepackage{soul, color} 
\soulregister\ref{7}  
\soulregister\cite{7} 
\renewcommand{\st}[1]{}

\begin{document}
\rmfamily

\title{2D Plasmonics for Enabling Novel Light-Matter Interactions}
\author{Nicholas Rivera$^{1\ast}$,Ido Kaminer$^{1\ast^\dagger}$,  Bo Zhen$^{2}$, John D. Joannopoulos$^{1}$ \& Marin Solja\v{c}i\'{c}$^{1}$}

\affiliation{$^{1}$Department of Physics, MIT, Cambridge, MA 02139, USA \\
$^{2}$Research Laboratory of Electronics, MIT, Cambridge, MA 02139, USA \\
$^\ast$These authors contributed equally to this work. \\
$\dagger$ Corresponding author e-mail: nrivera@mit.edu}

\noindent	

\maketitle


\noindent
\textbf{The physics of light-matter interactions is strongly constrained by both the small value of the fine-structure constant and the small size of the atom. Overcoming these limitations is a long-standing challenge. Recent theoretical and experimental breakthroughs have shown that two dimensional systems, such as graphene, can support strongly confined light in the form of plasmons. These 2D systems have a unique ability to squeeze the wavelength of light by over two orders of magnitude. Such high confinement requires a revisitation of the main assumptions of light-matter interactions. In this letter, we provide a general theory of light-matter interactions in 2D systems which support plasmons.  This theory reveals that conventionally forbidden light-matter interactions, such as: high-order multipolar transitions, two-plasmon spontaneous emission, and spin-flip transitions can occur on very short time-scales - comparable to those of conventionally fast transitions. Our findings enable new platforms for spectroscopy, sensing, broadband light generation, and a potential test-ground for non-perturbative quantum electrodynamics.}

\section{Introduction}
A fundamental process in light-matter interaction is spontaneous emission, in which an atom in an excited state can lower its energy by emitting a quanta of light \cite{dirac1927quantum}\footnote{This is to be distinguished from the more general phenomenon of spontaneous decay, in which an atom can decay by transferring energy into other channels, such as phonons. This is an important distinction which is discussed in [Ford, Weber 1984]. Radiative decay should include decay into surface plasmons, because in the absence of loss, these plasmons can transfer energy to infinity, just like a lossless far-field photon. On the other hand, any part of the decay proportional to $\text{ Im } \epsilon$ should be included as non-radiative decay because the energy transfer is into whatever provides losses within the dielectric, such as phonons and particle-hole excitations. These modes of energy transfer are notoriously difficult to detect, and thus it is preferable for the purposes of spectroscopy and other optical applications to have emission into propagating electromagnetic waves.}. In principle, spontaneous emission should occur in any atomic transition, yet in practice, the vast majority of transitions are “forbidden” by atomic selection rules\cite{cohen1992atom}. For example, rates of transitions that change the orbital angular momentum of the electron by n scale like $\alpha (ka_0)^{2n}$, where $\alpha \approx \frac{1}{137}$ is the fine-structure constant, $k$ is the wavenumber of the photon, and $a_0$ is the characteristic atomic size (e.g; Bohr radius). Because atoms are typically much smaller than the wavelength of the photon that they emit, the emission of multipole photons is a very slow process. In practice, electric octupole photon emission, magnetic quadrupole photon emission, and any higher order multipole process is too slow to be spectroscopically observable\cite{kramida2012nist}.  For a similar reason, transitions from singlet states to triplets states are considered impossible in light-matter interaction, to the degree that light inability to change the electron spin is considered a fundamental rule.

Another important class of slow atomic emission processes are those in which an atom emits two quanta of light. The rate of this process scales as $(\alpha(ka_0)^2)^2$\cite{breit1940metastability}. Due to the small size of the atom and the small value of the fine-structure constant, this emission of two quanta tends to be 8-10 orders of magnitude slower than the emission of one quantum of light. In fact, although two-photon spontaneous emission was predicted in the early 1930s, the first direct observation of two-photon spontaneous emission in Hydrogen occurred in 1996 \cite{cesar1996two}. While two-photon (and three-photon) absorption can be achieved with very high intensities of light, the only way to enhance spontaneous emission is by enhancing the electromagnetic field at the single-photon level. Enhancing both multipolar and multiplasmon processes would require unprecedented light confinement.

Here we show that with the recent advances of 2D plasmonics\cite{nagao2001dispersion,diaconescu2007low,liu2008plasmon,rugeramigabo2008experimental,jablan2009plasmonics,PhysRevLett.105.016801,grigorenko2012graphene,fei2012gate,jablan2013plasmons,tielrooij2015electrical}, it is possible to overcome all of the mentioned conventional limitations of light-matter interactions. Using a general theory of light-matter interactions, we show that the rates of high-order multipolar transitions, two-plasmon spontaneous emission, and spin-flip transitions become significantly enhanced to the point at which they compare to dipole transition rates and thus become easily observable. Applications of this work include novel platforms for spectroscopy,  a new wealth of data on the electronic structure of atoms and molecules, sensors, organic-light sources, entangled light generation, and supercontinuum generation.

\section{2D Plasmons and Light-Matter Interaction}


Plasmons are understood in the framework of classical electrodynamics as being coherent propagation of surface charge associated with tightly confined electromagnetic fields. The fact peculiar to 2D plasmons is that their wavelength, $\lambda_{pl}$, which is determined by the dispersion relation, can be hundreds of times shorter than the wavelength of a photon, $\lambda_0$, at the same frequency. This wavelength is shorter than the free-space wavelength by the confinement or squeezing factor, $\eta \equiv \frac{\lambda_{pl}}{\lambda_0}$. For instructional purposes, we present the simplest 2D plasmon model that accurately describes plasmons in doped graphene below the onset of interband transitions, with squeezing factor 
\begin{equation}
\eta \equiv \frac{1+\epsilon_r}{2\alpha}\frac{\hbar\omega}{2E_F},
\end{equation}
where $E_F$ is the Fermi energy, and $\epsilon_r$ is the permittivity of the substrate. In graphene plasmons, this squeezing factor has been predicted to be as high as 300\cite{jablan2009plasmonics}, with values of 220 having already been observed below the interband regime and values of 240 being observed in the interband regime\cite{liu2008plasmon}.  In other 2D plasmons such as silver, squeezing factors as high as 300 have been observed, corresponding to plasmon wavelengths of around 5 nm at photon wavelengths of $1.5 \mu m$\cite{nagao2001dispersion}.  In beryllium, acoustic plasmons have been observed at visible frequencies with plasmon wavelengths of merely 1 nm\cite{diaconescu2007low}.   

In our study, we look at the emission of 2D plasmons by Hydrogen-like and Helium-like atomic emitters in the vicinity of a 2D plasmonic system given by the Hamiltonian:
\begin{align}
&H = H_{atom} + H_{em}  \\
&H_{atom} = \left(\sum_i \frac{p_i^2}{2m_e} - \frac{Ze^2}{4\pi\epsilon_0r_i}\right) + H_{e-e} + H_{SO} \nonumber \\
&H_{em} = \sum_i \frac{e}{m_e}\mathbf{p}_i\cdot\mathbf{A}(\mathbf{r}_i) + \frac{e^2}{2m_e}\mathbf{A}^2(\mathbf{r}_i) + \frac{e}{m_e}\mathbf{S}_i\cdot\mathbf{B}(\mathbf{r}_i) \nonumber,
\end{align} 
where $\mathbf{A},\mathbf{B}$ are the vector potential and magnetic fields, $\mathbf{r}_i$ denotes the position of the $i$th electron, $\sigma_i$ denotes the spin of the $i$th electron, $H_{SO}$ is the spin-orbit coupling, and $H_{e-e}$ is the electron-electron interaction. The parameters of the plasmon and emitter relevant to our calculations are shown schematically in Figure 1. This Hamiltonian, with the appropriate field operators, is sufficient to describe radiative decay and nonradiative decay mediated by interactions of the electromagnetic field with orbital and spin degrees of freedom of the electron in an atom. It also describes intercombination transitions, multiplasmon emission, and other processes higher order in the perturbation theory. In this study, we examine all of these in order to provide a broad picture of atom-light interactions in plasmonics. Our calculations are discussed in greater detail in the Supplementary Materials.

\section{Singlet-Triplet Transitions}

The first type of transitions that we analyze here are those for which the initial state does not directly connect to the final state by the emission of light, but rather  couples to a virtual state whose symmetry is compatible with that of the final state. Take as an example a radiative transition between a spin-singlet state, S, and a spin-triplet state, T. While this transition can happen directly through the magnetic field, these magnetic transitions are slow, happening on the time scale of milliseconds for magnetic dipole transitions, hundreds of seconds for magnetic quadrupole transitions, and so on. The dominant process for systems with large spin-orbit coupling is a second-order process in which an initial triplet state connects to a virtual singlet state, $S_n$, through the spin-orbit coupling, and then the virtual state connects to the final singlet state and emits light. The decay rate is then given by the second-order perturbation theory expression: $
\Gamma(T\rightarrow S) = \frac{2\pi}{\hbar^2}\sum\limits_{\mathbf{q}} \Big| \sum_n \frac{\langle S,\mathbf{q} |H_{em}|S_n,0 \rangle \langle S_n,0|H_{SO}|T \rangle}{E_{T}-E_{S_{n}}} \Big|^2\delta(\omega_{\mathbf{q}}-\omega_0),
$ where $\omega_0$ is the transition frequency and $\mathbf{q}$ is the wavevector of an emitted plasmon.

Despite this being the dominant process driving the singlet-triplet transition, it is still generally slow, and much effort has been given to speeding it up. One motivation for speeding up such a transition is that its slowness impedes the progress of continuous-wave organic dye lasing. In organic dye lasers, one would like to achieve population inversion between a ground and excited singlet state. However, what can happen is that a relatively rapid intersystem crossing takes place, moving the singlet population to a triplet state. Because of the generic slowness of the singlet-triplet transition, the population of the triplet state gets stuck and therefore continuous-wave lasing cannot occur. Efforts to speed up this transition have focused on enhancing the spin-orbit coupling, and reducing the rate of intersystem crossings. However, as we shall show here, it is possible to combine enhanced spin-orbit coupling with plasmonics in order to get very high singlet-triplet transition rates. 

To calculate the decay rate using second-order perturbation theory, one should sum over all possible intermediate states. However, in the cases of interest, there is an intermediate singlet state that is nearly degenerate with the triplet state, which then gives the dominant contribution to the transition. The advantage of this approximation is that the spin-orbit and electromagnetic enhancements decouple. The Purcell factor, defined as the emission rate into plasmons divided by the emission rate into free-space photons, in the absence of plasmon losses is proportional to
\begin{equation}
F_p(T\rightarrow S) \sim \eta_0^3e^{-2\eta_0kz_0},
\end{equation}
where $\eta_0$ is the squeezing factor at the transition frequency $\omega_0$. $z_0$ is the separation between the atomic nucleus and the surface. This simple formula is only weakly modified by plasmonic losses, which the subsequent calculations take into account (see SI for further details).  In Figure 2, we illustrate the Purcell factors achievable for an emitter on top of graphene and monolayer silver as a function of plasmon squeezing. We show the Purcell factors at distances of 1, 2, and 5 nm away for two different plasmon quality factors (Q=10 and Q=100). Figure 2 demonstrates how 2D plasmons can enhance singlet to triplet transition by up to 7 orders of magnitude for realistic and experimentally readily available parameters. This enhancement is in addition to enhancements coming from increasing spin-orbit coupling. These high enhancement factors are in agreement with enhancements of electric dipole emitters on top of graphene \cite{koppens2011graphene}. At higher plasmonic losses, nonradiative decay rates increase, leading to the observation that losses counteract the effects of placing the emitter farther from the surface. Additional non-radiative decay is not problematic in applications where the goal is to quench the triplet population.

To conclude this section, we point out that the our approach allows calculation of the enhancement factor without
knowing anything about the spin-orbit coupling or even the structure of the atom/molecule. As we show below, there is another method of enhancing singlet-triplet transitions, which is through a direct emission into a plasmon. Such a transition is driven by the magnetic field of the plasmon.



\section{Multipolar Transitions}

In this section, we focus on multipolar transitions. Electric quadrupole and magnetic dipole transitions, which are the second fastest types of transitions, are fairly slow, but can be observed through conventional spectroscopic approaches. Significant theoretical and experimental efforts have gone towards speeding up these two processes \cite{zurita2002multipolar,zurita2002multipolar2,filter2012controlling,takase2013selection,yannopapas2015giant}. As mentioned previously, electric octupole, magnetic quadrupole, and higher order multipolar transitions are far too slow to observe. However, as we now demonstrate, due to the high squeezings achievable in graphene and other 2D plasmons, the contributions to radiative decay from very high-order multipolar transitions can become comparable to those from dipole transitions. In fact, transitions whose lifetimes approach the age of the universe can be brought down to lifetimes of hundreds of nanoseconds, corresponding to rate enhancements  (Purcell factors) of nearly $10^{24}$. Such rates are comparable to dipole transition rates in free space and therefore should be straightforwardly accessible through absorption spectroscopy.

To explain this, we calculate the rates of transitions where the electron orbital angular momentum changes by $n$ (En transition). The decay rate $\Gamma_n$ into plasmons scales with $\eta_0$ as:
\begin{equation}
\Gamma_n \sim \eta_0^{3+2(n-1)}e^{-2\eta_0k z_0},
\end{equation}
which makes it very clear that higher electric multipole transitions are enhanced by successively larger amounts. For achievable squeezing factors of 100, when n increases by 1, the enhancement increases by a factor of roughly 10000.  Therefore, one should expect that if the dipole transitions (n=1) are enhanced by $10^6$, then an E2, E3, E4, and E5 transition should be enhanced by $10^{10},10^{14},10^{18},$and $ 10^{22},$ respectively. Such an intuition is confirmed in Figure 3, where we plot the exact transition rates, with and without plasmonic losses, for the series of transitions $6\{p,d,f,g,h\} \rightarrow 4s$. In our simulations, the emitter is kept 5 nm away from the surface \footnote{The $6s \rightarrow 4s$ rate is zero when the atom is well separated from the surface due to the transversality of the electromagnetic field ($\nabla \cdot \mathbf{E} = 0$). The dominant decay mechanism for this process will be two-plasmon emission, which we compute in Section 5.}. In Figure 3(a), we plot the rates of radiative transitions in the Hydrogen $6 \rightarrow 4$ series relative to the rate of the dipole transition as a function of squeezing. This relative rate is independent of atom-surface separation. To give a particular example, conventionally the E5 transition is separated from the E1 transition by 22 orders of magnitude. The free-space rate of the E5 transition is around $10^{-16}$ sec$^{-1}$, corresponding to a lifetime which is two orders of magnitude less than the age of the observable universe. For high squeezing ($\eta_0 \sim 200$), the rate becomes $10^{7}$ sec$^{-1}$, which is an order of magnitude faster than the free-space E1 transition. This corresponds to a Purcell factor of roughly $10^{23}$. For extremely high squeezing $\eta_0 > 300$, the entire transition series lies within two orders of magnitude. 

In addition to radiation into plasmons, an excited emitter can decay non-radiatively into the conductor. These non-radiative decay channels can include, but are not limited to: phonons, particle-hole excitations, and impurities. These are incorporated in our calculations through the imaginary part of the permittivity of the surface. In Figure 3(b), we plot the total rates of the transitions considered in Figure 3(a), i.e; radiative+non-radiative (solid lines) in addition to the radiative decay rates (dashed lines). For low squeezing, the non-radiative decay is dominant. This reflects the well-known fact that non-radiative energy transfer strongly depends on the ratio $\frac{1}{kz_0}$. What is more valuable for optical applications is the situation at high squeezing $\eta_0 > 150$. There, the radiation into plasmons is dominant \footnote{For applications in which fast non-radiative decay is desired, we note that for considerably low wavelengths $\frac{1}{kz_0} \approx 1$, the non-radiative decay is much faster than one would naively expect. This is because the non-radiative decay still couples to the multipole moment relevant to the atomic transition. Therefore, the strong enhancement of rates coming from matching the atomic size to the wavelength of light compensates for the suppression of rates coming from a higher $kz_0$. In fact, the total rate - even when non-radiative energy transfer is dominant - increases as the wavelength shrinks. From this, we conclude that for both radiative and non-radiative decay, the high squeezing of 2D plasmons helps overcome the small size of the atom.}.

\section{Two-Plasmon Spontaneous Emission}

The above enhancements do not apply to $S \rightarrow S$ transitions, due to Gauss's law. In Hydrogen-like atoms, the dominant decay mode for $S \rightarrow S$ transitions will be by the emission of two light quanta, just as in free space. In two-plasmon spontaneous emission, the atom emits light in a range of frequencies between zero and the transition frequency $\omega_0$, consistent with energy conservation - the sum of the frequencies of emitted quanta equals the transition frequency.  This broad distribution of frequencies corresponds to a differential decay rate $d\Gamma/d\omega$.

The emission of two excitations of a field (photons or plasmons) is a second-order effect in perturbation theory. If the dimensionless coupling constant of atomic QED, $g$, is small, then two-quanta emission will be very slow compared to emission of a single field excitation. A good estimate for the coupling constant is $g^2 = \frac{\Gamma}{\omega_0}$. When this becomes nearly one, we expect the Fermi Golden Rule to fail because the width of the resulting atomic resonance compares to the frequency scale of variation of the continuum matrix elements\cite{cohen1992atom}. Taking this as our dimensionless coupling constant, we see that it scales as $g_n\sim\sqrt{\alpha(ka_0)^2\eta_0^{3+2(n-1)}e^{-\frac{4\pi\eta_0z_0}{\lambda_{ph}}}}$. 

For dipole transitions, this coupling constant suggests that the emission rate of two plasmons scales as $g^4 \sim \eta_0^6$. Our exact derivation shows this is indeed the case (see SI for further details).  We arrive at the following analytical expression for the enhancement of the differential decay rate (in the lossless limit) for any two-plasmon $S\rightarrow S$ transition in a Hydrogenic atom:
\begin{equation}
\frac{d\Gamma/du\Big|_{2~pl}}{d\Gamma/du\Big|_{2~ph}} = \frac{d\Gamma/d\omega\Big|_{2~pl}}{d\Gamma/d\omega\Big|_{2~ph}} = 72\pi^2\left(\eta_0^3e^{-2\eta_0kz_0}\right)^2\times(u-u^2)^3e^{8\eta_0kz_0u(1-u)},
\end{equation}
where $u = \frac{\omega}{\omega_0}$, and $\eta_0$ is the squeezing factor at the spectral peak of the emission. Said peak occurs for $\omega = \frac{1}{2}\omega_0$, or equivalently, u=1/2. $\frac{d\Gamma}{d\omega}(\omega')$ corresponds to the rate of emission of two plasmons per unit frequency in which one plasmon is at frequency $\omega'$ and the other is at frequency $\omega_0 - \omega'$.

In Figure 4(a), we plot the distribution of emitted plasmons as a function of normalized frequency, $u$. The spectral distribution narrows as a function of atom-surface separation, $z_0$, which we illustrate in Figure 4(b). This arises from the $u$ dependence of the exponential in Equation (5). What this shows is that through precise control of the position of an emitter above a surface supporting plasmons, not only can the rate of emission be tuned, but also the spectrum of emission.

In Figure 4(c), we plot an estimate of the total decay rate for the transition $4s \rightarrow 3s$, which is computed by summing over 10 virtual discrete states (our estimate is discussed in the Supplementary Materials). The rate converges sufficiently for 10 virtual states. Even for modest squeezings of around 100, and at separations of 1 nm, the two-plasmon spontaneous emission rate exceeds 1 ns$^{-1}$, in stark contrast with the typical rate of roughly 1 min$^{-1}$ in free-space. For squeezings beyond 200, it is possible to get emission rates exceeding 1 ps$^{-1}.$ We compare the rate of this transition with that of the $4s \rightarrow 3p$ single-plasmon dipole transition in Figure 4(d). In the region of extreme squeezing (200-500), the two-plasmon emission is only 1-2 orders of magnitude slower. 

We also observe that the ratio of the rates has very weak distance dependence. This arises from the fact that at greater atom-surface separations, the distribution of emitted plasmons narrows. Therefore, most of the emitted plasmons have frequencies near $u=1/2$. When this happens, the distance dependence of the two-plasmon decay rate approaches $e^{-2\eta_0 kz_0}$, which is the same distance dependence as that of the one-plasmon decay rate. This is a feature of the $\sqrt{q}$ dispersion of plasmons well-described by the Drude model. For plasmons with linear dispersion (such as acoustic plasmons), the distance dependence of the two-plasmon differential emission rate is simply $e^{-4\eta_0kz_0}$ (i.e; no $u$ dependence). 

The results summarized in Figures 4(c) and 4(d) establish two conclusions. The first is that two-plasmon spontaneous emission can compete with one-plasmon spontaneous emission, meaning that the multi-quanta nature of spontaneous emission can no longer be neglected.. The second is that for achievably high squeezings, high-order effects in QED become observable. This signals the onset of non-perturbative quantum electrodynamics.

\section{Summary and Outlook}
2D plasmonics presents a unique opportunity to access radiative transitions that were considered inaccessible. Multi-plasmon processes, spin-flip radiation, and very high-order multipolar processes have been shown to be competitive with the fastest atomic transitions in free-space. Our findings have been summarized in Table I. These findings pave the way to using and controlling new light-matter interactions. Moreover, because 2D plasmonics bridges the gap between light and the smallest emitters, it is clear that for high squeezing, the dipole approximation is completely in doubt for larger emitters such as molecules, quantum dots, and Bloch electrons.

2D plasmons on planar surfaces present a number of attractive features such as large squeezing over a wide range of frequencies, as well as large interaction areas thus facilitating control of light-matter interaction without the usual constraints of small cavities and very narrow structures. Unlike the case of plasmonic waveguides, where propagation distance and thus losses are crucial, in 2D systems for light-matter interaction applications, losses can be overcome by excitation of plasmons over a large area. Such excitation has been done with array nanoribbons. In particular cases when one is interested in the strongest possible signal, achieving such large area plasmon enhancement is especially important, because unlike localized plasmonic resonances, it allows a larger quantity of emitters to be studied at once. Another attractive feature of this approach is modularity - the fact that unlike systems in which strong electromagnetic interaction with matter necessitates large emitters, low emission frequencies, and/or mostly non-radiative decay\cite{andersen2011strongly,rukhlenko2009spontaneous,jain2012near}, our approach works for any atomic size and at a wide range of frequencies between visible and long-IR.


With regards to applications of this work to fundamental light-matter interactions, we believe that a number of fruitful extensions are within reach. Beyond the processes that we considered in this work, one can consider combinations of these processes, such as:
\begin{enumerate}
\item{Multi-plasmon transitions mediated by higher-order multipole virtual transitions. For example, a two-plasmon emission with a total angular momentum change of 4 for the electron, by way of an intermediate quadrupole virtual transition.}
\item{Three and higher order plasmon emission and absorption processes}
\item{A second order absorption process in which a plasmon and a far-field photon are absorbed, leading to large changes in energy and angular momentum of the electron, due to the photon and plasmon respectively.}
\end{enumerate}
All of these transitions should be significantly enhanced using 2D plasmonics, but out of reach of conventional plasmonics and photonics, without extremely high intensities. We note that similar considerations to those in this letter yield qualitatively similar conclusions for 1D plasmons. Although highly confined 1D plasmons have been observed in ultrathin wires of Au and other materials \cite{nagao2007sound,rugeramigabo2010one}, a potential difficulty in exploiting 1D plasmonics is low interaction area between the emitter and the wire. Another difficulty is that for radii much larger than that of the atom, the enhancements will be several orders of magnitude less. However, in Ref. 25, radii as low as 3 nm have been achieved, meaning that it should be possible to obtain decay rates comparable to those obtained in this letter. Such a possibility has not been explored. 

The potential applications of this work include: spectroscopy for inferring electronic transitions which cannot be determined with photons, sensors based on forbidden transitions, organic-light sources arising from fast singlet-triplet transitions, fast entangled light generation, and fast generation of broadband light with tunable width in the visible or IR.

\section{Acknowledgements}
This work was partly supported by the Army Research Office through the Institute for Soldier Nanotechnologies
under contract no W911NF-13-D-0001. M.S. (analysis and reading of the manuscript) was supported by S3TEC, an Energy
Frontier Research Center funded by the US Department of Energy under grant no. DE-SC0001299. I.K. was supported in
part by Marie Curie grant no. 328853-MC-BSiCS.
\clearpage
\begin{figure}[t]
\includegraphics[width=150mm]{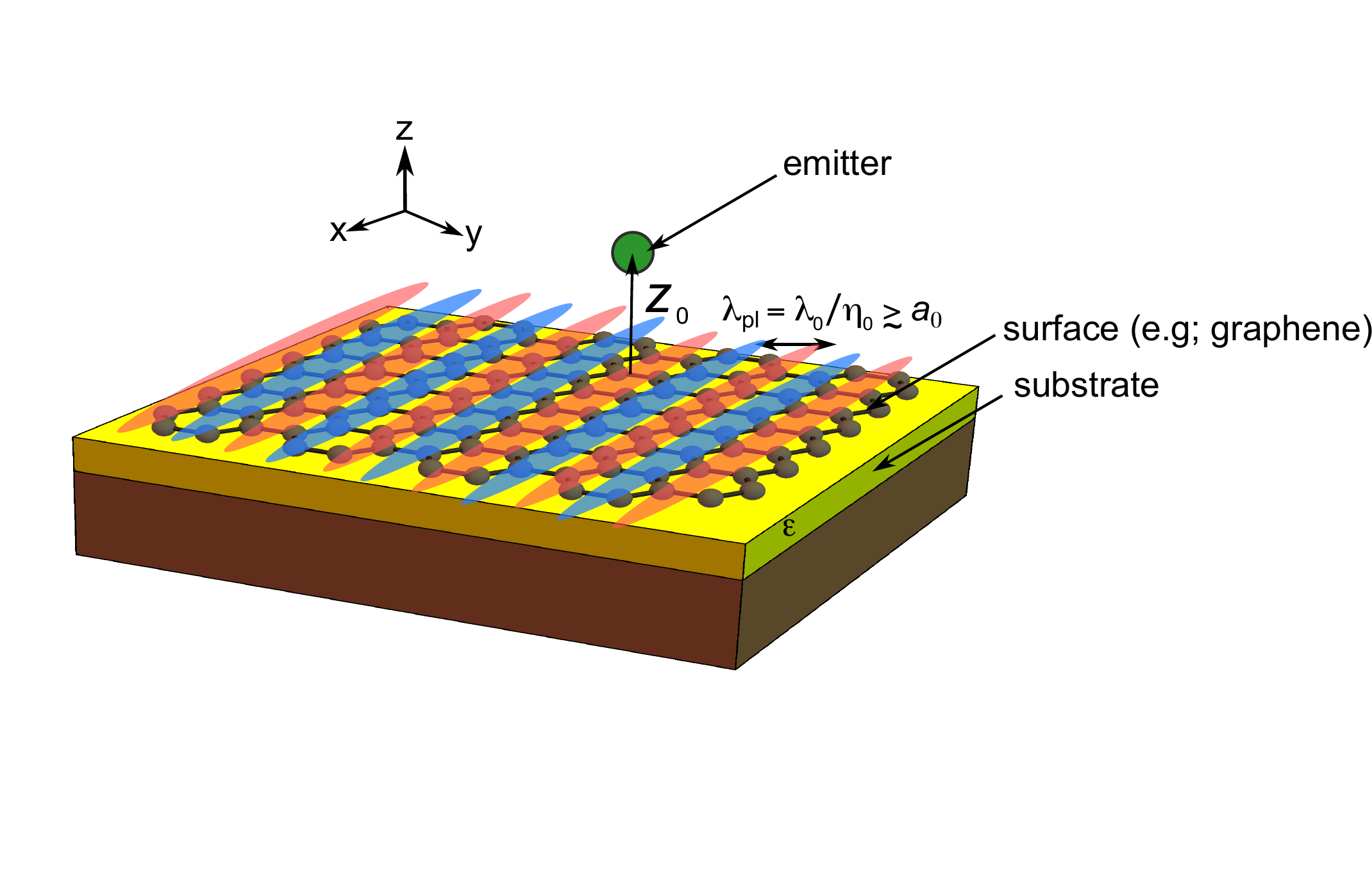}
\caption{\textbf{2D Plasmonics.} A schematic of an emitter (not-necessarily dipolar) above a 2D material of conductivity $\sigma_s$ supporting plasmons with wavelength far shorter than the photon wavelength, and approaching the atomic size.}
\end{figure}
\begin{figure}[h]
\includegraphics[width=180mm]{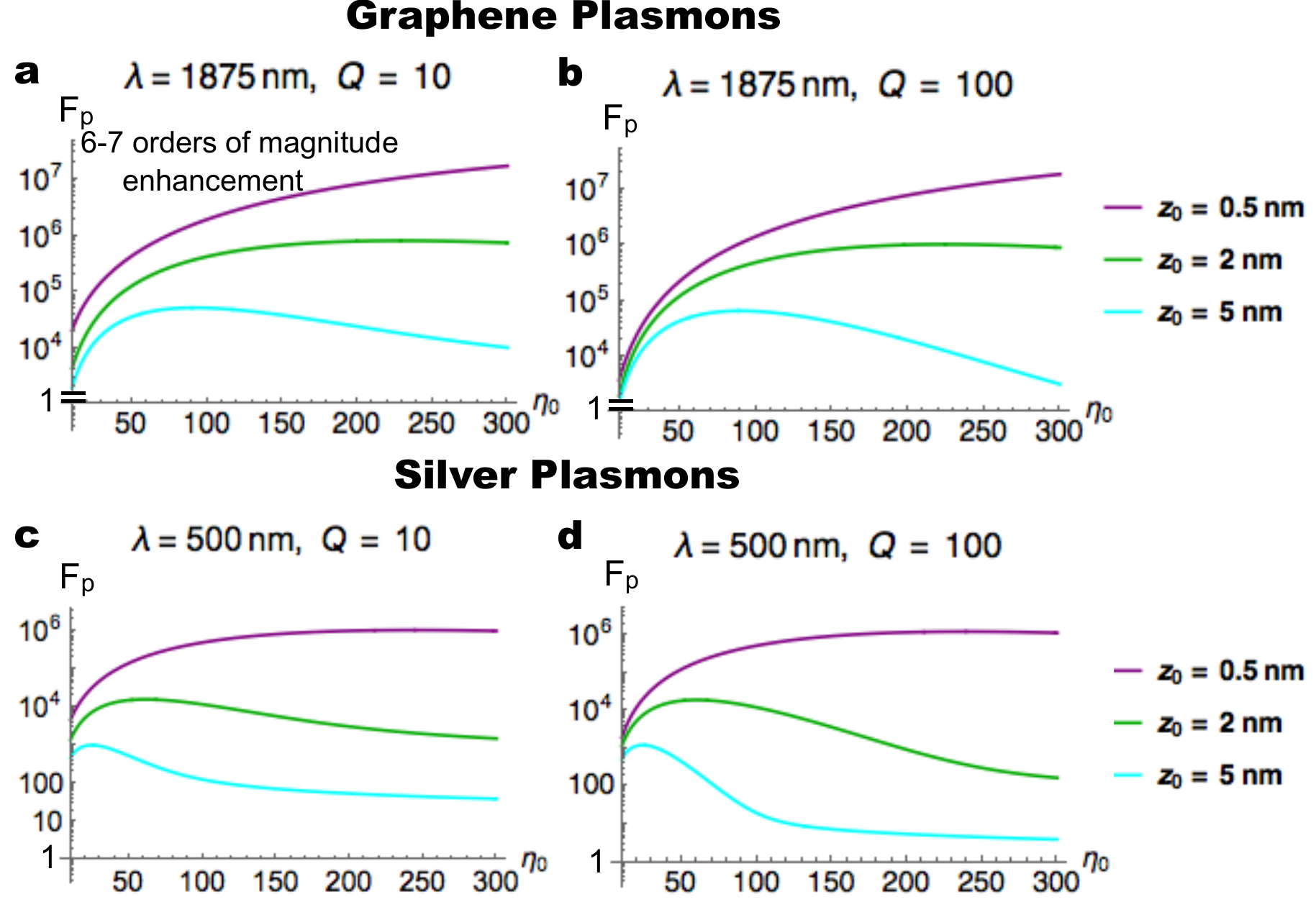}
\caption{\textbf{Enabling singlet to triplet transitions.} Purcell factors for a dipolar singlet-triplet transition as a function of squeezing factor $\eta_0$ for (a,b) plasmons in graphene and (c,d) plasmons in 2D silver,  for different atom-plane separations, $z_0$ and quality factors, $Q$. These results apply to any intercombination transition.}
\end{figure}
\clearpage
\begin{figure}[h]
\includegraphics[width=180mm]{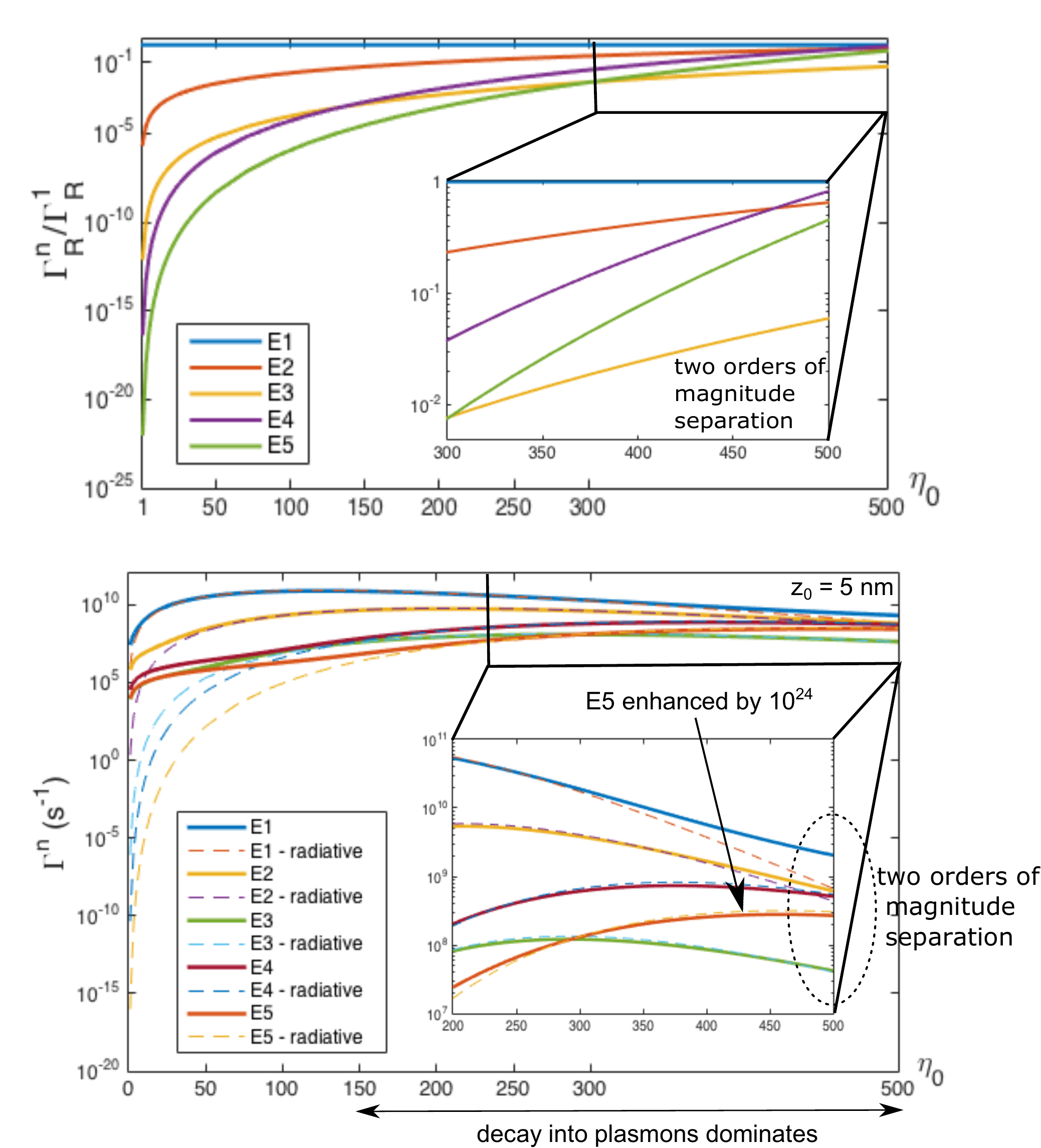}
\caption{\textbf{Convergence of the Multipoles.} (a) Rates of radiation into surface plasmons for various multipole transitions in Hydrogen. The transition series considered here is $6\{p,d,f,g,h\}\rightarrow 4s$. The emitter is situated 5 nm above the surface of graphene. (b). Total rates (radiative + non-radiative) of decay for the same transition series in Hydrogen. Dashed lines show radiative rates, which agree with the total rates at high squeezing.}
\end{figure}
\begin{figure}[h]
\centering
\includegraphics[width=185mm]{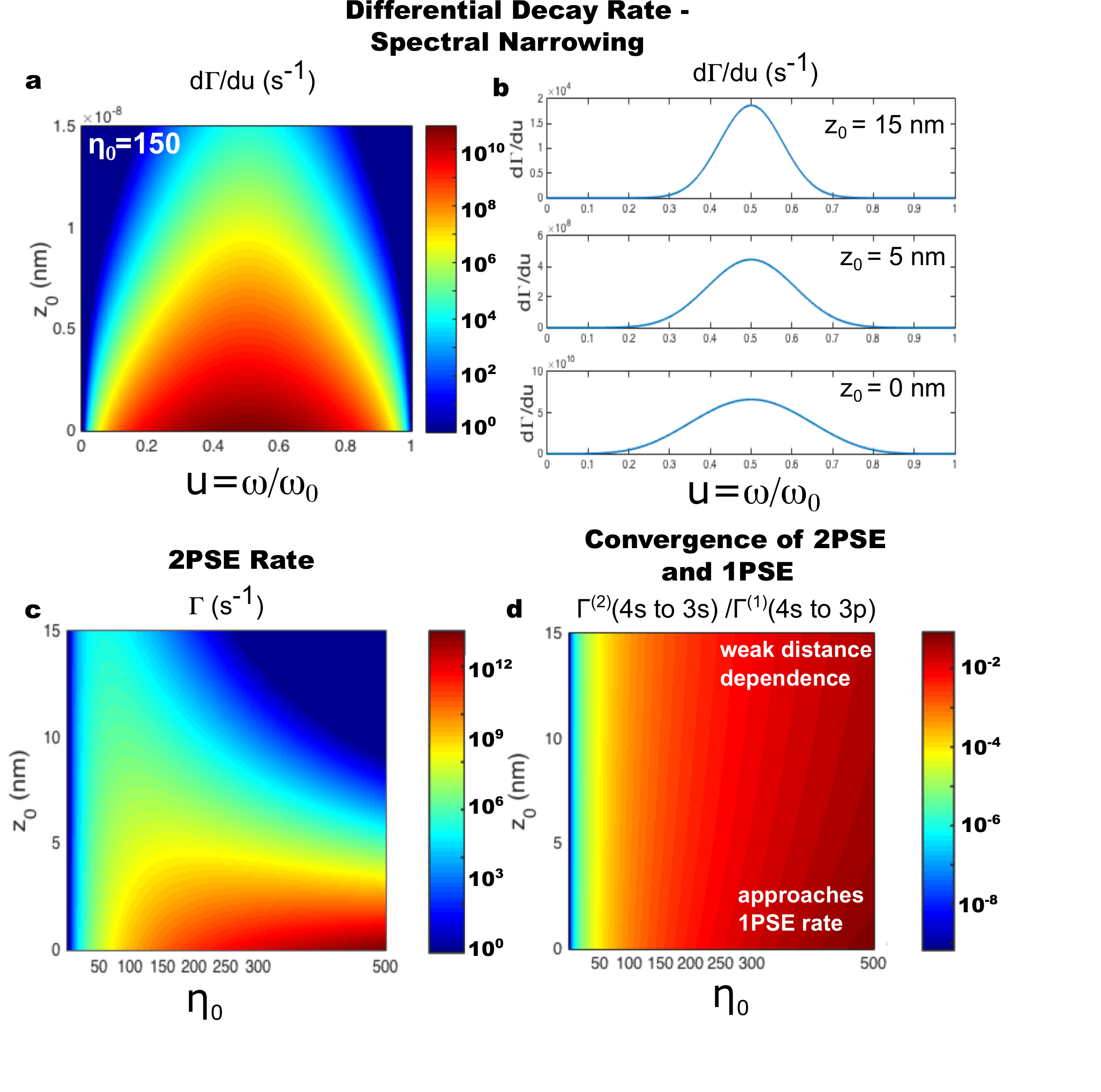}
\caption{\textbf{Enabling Two-Plasmon Radiative Energy Transfer.}  (a) Emission spectrum of two-plasmon spontaneous emission (2PSE) as a function of frequency and atom-surface separation for the Hydrogen $4s \rightarrow 3s$ transition above Graphene. $\eta_0$ is chosen to be 150. (b). Linecuts of (a) for atom-surface separations of 0, 5, and 15 nm. (c) Total decay rate (in sec$^{-1}$) for this transition as a function of squeezing and atom-surface separation. (d) Comparison of two-plasmon emission rate to the emission rate for a single plasmon $4s \rightarrow 3p$ transition as a function of squeezing and separation. }
\end{figure}
\clearpage
\begin{table*}[t]
\centering
\begin{tabular}{l|l|l|c}
\hline
Transition & Free Space $\Gamma$ &2D Plasmon $\Gamma$ & $\eta$ Enhancement   \\
\hline
E1 & $\alpha\left(\frac{\hbar \omega}{m_ec^2}\right) $ & $\alpha \left(\frac{\hbar \omega}{m_ec^2}\right)\eta^3$ &3 \\
En\footnote{Losses can change these significantly at low squeezing or short separation.} &$\alpha\left(\frac{\hbar \omega}{m_ec^2}\right)(ka_0)^{2(n-1)}$  & $\alpha\left(\frac{\hbar \omega}{m_ec^2}\right)(ka_0)^{2(n-1)} \eta^{3+2(n-1)}$& 3+2(n-1)\\
Spin-Flip Ek\footnote{Not-including the spin-orbit matrix element, which approximately cancels in the Purcell enhancement.} & $\alpha\left(\frac{\hbar \omega}{m_ec^2}\right)(ka_0)^{2(n-1)} $ &$ \alpha\left(\frac{\hbar \omega}{m_ec^2}\right)(ka_0)^{2(n-1)}\eta^{3+2(n-1)}$ & 3+2(n-1) \\
2PSE\footnote{2PSE = Two-plasmon spontaneous emission} (Dipole) &$\alpha^2(ka_0)^4$  & $\alpha^2(ka_0)^4\eta^6$ & 6\\
\hline
M1\footnote{M1 and Mn are discussed in the Supplementary Materials} & $\alpha\left(\frac{\hbar \omega}{m_ec^2}\right)^2 $ & $\alpha\left(\frac{\hbar \omega}{m_ec^2}\right)^2 \eta$ & 1 \\
Mn$^{a}$ &$\alpha\left(\frac{\hbar \omega}{m_ec^2}\right)^2(ka_0)^{2(n-1)}$  & $\alpha\left(\frac{\hbar \omega}{m_ec^2}\right)^2(ka_0)^{2(n-1)} \eta^{2(n-1)+1}$ &1+2(n-1)\\
\end{tabular}
\caption{A summary of derived scalings of rates and rate enhancements for emitters on top of lossless 2D plasmons at zero displacement from the plasmon supporting surface. The rates for an emitter at finite distance from the plasmon are suppressed by $e^{-2\eta_0 kz_0}$, which is of order unity for an emitter within a reduced plasmon wavelength of the surface.}
\end{table*}
\clearpage

\bibliographystyle{unsrt}
\bibliography{2DPlasmonsBib}


\end{document}